\documentclass[12pt]{article}
\usepackage{bbm}
\usepackage{amsmath}
\usepackage{amssymb}
\usepackage{amstext}
\usepackage{amsfonts}
\usepackage{graphicx,epsfig}
\usepackage{indentfirst}

\newcommand{\stl}{\scriptstyle}
\newcommand{\mbd}{\mathbf}

\newtheorem{theorem}{Theorem}[section]
\newtheorem{lemma}[theorem]{Lemma}
\newtheorem{proposition}[theorem]{Proposition}

\begin{document}
\baselineskip 20pt
\title{Classification of the Entangled States $2 \times M \times N$}
\author{Jun-Li Li$^{a}$\; ~and
Cong-Feng Qiao$^{a,b}$\footnote{Corresponding author.}\\[0.5cm]
{\small $a)$ Dept. of Physics, Graduate
University, the Chinese Academy of Sciences}  \\
{\small YuQuan Road 19A, 100049, Beijing, China}\\
{\small $b)$ Theoretical Physics Center for Science Facilities
(TPCSF), CAS}\\
{\small YuQuan Road 19B, 100049, Beijing, China}\\
}
\date{}
\maketitle

\begin{abstract}
We extend the matrix decomposition method(MDM) in classifying the
$2\times N\times N$ truly entangled states to $2\times M\times N$
system under the condition of stochastic local operations and
classical communication. It is found that the MDM is quite practical
and convenient in operation for the asymmetrical tripartite states,
and an explicit example of the classification of $2\times 6\times 7$
quantum system is presented.
\end{abstract}

\section{Introduction}

Entanglement is an essential feature of quantum theory, describing a
quantum correlation that exhibits nonlocal properties. In the
seminal work \cite{EPR}, Einstein, Podolsky, and Rosen (EPR)
demonstrated through a Gedanken experiment that the quantum
mechanics (QM) can not provide a complete description of the
``physical reality" for two spatially separated but quantum
mechanically correlated particles state which is now known as
entangled state. The subsequent Bell theorem manifest the nonlocal
character of the quantum correlation in the violation of Bell's
inequalities \cite{Bell}. As the quantum information science
develops, the impact of entanglement goes far beyond the testing of
the conceptual foundations of QM. Entanglement is now of central
importance in the quantum information theory (QIT) and is thought as
the key physical resource to realize quantum information tasks, such
as quantum cryptography \cite{cryp-1, cryp-2}, superdense coding
\cite{superdense-1, superdense-2}, and quantum computation
\cite{quantum-computation}, etc.  This necessitates the qualitative
and quantitative description of the entanglement \cite{quan-enta}.
However due to the lack of suitable tools for characterizing the
entanglement, very limited quantum state space was explored in the
quantum information theory.

In quantum information processing (QIP), two states are suited to
implement the same task if they can be mutually converted by
stochastic local operations and classical communication (SLOCC)
\cite{three-two}, and therefore they are said to be in the same
equivalent class. For three qubits, known result is that there are
two kinds of true tripartite entanglement classes for pure state,
namely, GHZ and W states \cite{three-two}. As the dimensions of each
party increases nontrivial aspect shows up, i.e., non-local
parameters may resides in the entangled states of $2\times N\times
N$ system when $N\geq 4$ \cite{range-244,class-nn}. Many
investigations concerned the classifications of $2\times M\times N$
states has been done in \cite{ range-244, range-2mn,
tripartite-qubit}. In the Refs.\cite{range-244, range-2mn}, an
iterated method was introduced to determine all the inequivalent
classes of the entangled states of $2\times M\times N$ system based
on the ``range criterion", where the entanglement classification of
the low dimension system is a prerequisite for the high dimensions
ones. Practical classifications of dimensions up to $2\times 4\times
4$ and the related systems of $2\times (M+4)\times (2M+4)$ were
given in \cite{range-244}. With the increasing of dimensions, the
complexity of the method grows dramatically because of the iterated
nature of their inequivalent proof of the entanglement classes. In a
recent work \cite{class-nn} a novel method of classifying the pure
state of $2\times N\times N$ systems was introduced in which all the
inequivalent true tripartite entanglement classes can be determined
directly by using merely the elementary operations on the cubic grid
form of the state.

The present work deals with the more general case: quantum state of
$2\times M\times N$ systems (pure state if not specified). We show
that the method we introduced in \cite{class-nn} can be generalized
to the classification of true entangled states of $2\times M\times
N$ systems. Although the main tools are the same for $2\times M
\times N$ with that of $2\times N\times N$, the generalization is
nontrivial and the method for $2\times M \times N$ can help the
general classification of $L\times M\times N$ systems. All the
inequivalent classes can be generated directly and no followed-up
inequivalence proof of these classes is needed. The content goes as
follows, in section 2, by representing the $2\times M\times N$ state
in the form of matrix pairs, the $2\times M\times N$ states are
divided into inequivalent sets under SLOCC. The detailed
classification procedures with these inequivalent sets are presented
in section 3 and a concrete example of classification of $2 \times
6\times 7$ system is given. Finally, in section 4 we give some
concluding remarks.

\section{ Matrix pair representation of $2\times M \times N$ state }

\begin{figure}\centering \scalebox{0.8}{\includegraphics{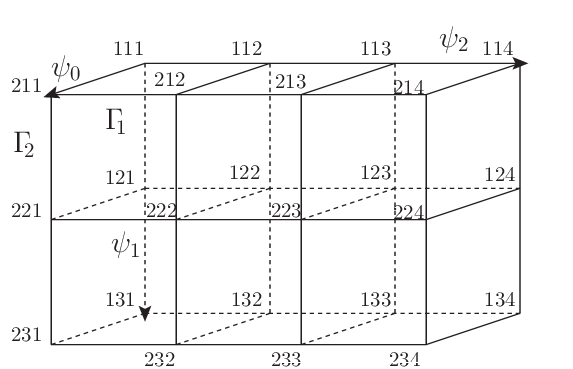}}
\caption{\small The cubic form for $2\times 3\times 4$ state. The
node in the grid labeled $ijk$ represents the matrix elements
$\Gamma_{\{i,j,k\}}$.} \label{pic-cube-234}
\end{figure}

Adopt the conventions of \cite{class-nn}, an arbitrary state of
$2\times M\times N$ can be written as
\begin{eqnarray}
|\Psi_{2\times M\times N}\rangle =  \sum_{i,j,k} \Gamma_{\{i,j,k\}}
\, |i\rangle_{\psi_0} |j\rangle_{\psi_1} |k\rangle_{\psi_2}
\label{psi1psi2}\; ,
\end{eqnarray}
where, $\psi_0$ represents the first qubit, $\psi_{1}$ and
$\psi_{2}$ has the dimension of $M$ and $N$ separately;
$\Gamma_{\{1,j,k\}}$ and $\Gamma_{\{2,j,k\}}$ are $M\times N$
complex matrices (we assume $M\leq N$ without loss of generalities).
Then the state can be written in the following compact form
\begin{eqnarray}
|\Psi_{2\times M\times N}\rangle =
\begin{pmatrix}
\Gamma_{\!1} \\
\Gamma_{\!2} \\
\end{pmatrix} . \label{2mn-mat-pair}
\end{eqnarray}
Clearly, to every state of $2\times M\times N$, there is a form of
Eq.(\ref{2mn-mat-pair}) that corresponds to it, and a pictorial
description of the state is straightforward, see
Fig.(\ref{pic-cube-234}).

The reduced density matrix of state $\Psi_{2\times M\times N}$ is
defined as $\rho_{\psi_i} = \mathrm{Tr}_{\neg\,\psi_{ i}}
[|\Psi\rangle \langle \Psi|]$, where $i\in \{0,1,2\}$. For
three-partite systems, true (or genuine \cite{three-two})
entanglement means that reduced density matrices of each partite
have full ranks. Let $r$ denote the rank of matrix hereafter, then
$r(\rho_{\psi_0}) = 2, r(\rho_{\psi_1}) = M, r(\rho_{\psi_2}) = N$
for the true entangled state of $2\times M\times N$ systems. The
density matrix in the form of the matrix pairs can be expressed as
\begin{eqnarray}
\rho_{\psi_0, \psi_1, \psi_2} = (\Gamma_{i})_{jk}
(\Gamma_{i'})_{j'k'}^{*} \; ,
\end{eqnarray}
where $i,i' = 1,2; j,j' = 1, 2, \cdots, M; k,k' = 1, 2, \cdots, N$.
The reduced density matrix (take $\psi_2$ as an example) then is
\begin{eqnarray}
\rho_{\psi_{2}} & = & \mathrm{Tr}_{\psi_0,\psi_1} ( \rho_{\psi_0,\psi_1,\psi_2}) \nonumber \\
& = & \sum_{ij} (\Gamma_{i})^{*}_{jk'}(\Gamma_{i})_{jk} \nonumber \\
& = & \sum_{i} \Gamma_{i}^{\dagger}\Gamma_{i} \; .
\label{density-cube}
\end{eqnarray}

\begin{lemma}
$\forall$ $i\in \{0,1,2\}$, $\mathrm{Det}(\rho_{\psi_i})=0$, if and
only if the cubic form of the three-partite state (see
Fig.(\ref{pic-cube-234})) can be transformed into a form where at
least one plane perpendicular to axis $\psi_i$ are zero planes
(plane with all its coefficients are zeroes) via ILOs.
\label{lemma1}
\end{lemma}
The proof is presented in Appendix \ref{appdixlemma1}. Thus if
$\mathrm{Det}(\rho_{\psi_i})=0, i\in \{0,1,2\}$, the entanglement of
$2\times M\times N$ system reduces to the case of $2\times M' \times
N'$ with $M'<M$ or/and $N' < N$ which should in principle be
considered as an entanglement system of $2\times M' \times N'$.

\section{Classification of $2\times M\times N$ State}

Two $2\times M\times N$ states $\widetilde{\Psi}$ and $\Psi$ are
said to be SLOCC equivalent if they are connected via invertible
local operators (ILOs). That is, $\widetilde{\Psi}$ is SLOCC
equivalent to $\Psi$ if
\begin{eqnarray}
|\widetilde{\Psi}_{2\times M\times N}\rangle = T \otimes P \otimes
Q\, |\Psi_{2\times M\times N}\rangle \; , \label{2mn-general-ILO}
\end{eqnarray}
where $T, P, Q$ are invertible complex matrices of dimension
$2\times 2$, $M\times M$, and $N\times N$ which act on $\psi_0$,
$\psi_{1}$, $\psi_{2}$, respectively. Neglecting the extra factor of
the determinant of matrices, $T$, $P$, and $Q$ correspond to the
special linear groups of $SL(2,\mathbb{C}), SL(M,\mathbb{C}),
SL(N,\mathbb{C})$ \cite{three-two}. Takes the wave function
$|\Psi_{2\times M\times N}\rangle$ in the matrix pair form [i.e.,
Eq.(\ref{2mn-mat-pair})], the ILO operators $T$, $P$, $Q$ in
Eq.(\ref{2mn-general-ILO}) take the following form
\begin{eqnarray}
|\widetilde{\Psi}_{2\times M\times N}\rangle & = & \begin{pmatrix}
    t_{11} & t_{12} \\
    t_{21} & t_{22} \end{pmatrix} \begin{pmatrix}
    P \Gamma_{\!1} Q \\
    P \Gamma_{\!2} Q \end{pmatrix}, \label{2mn-mat-pair-ILO}
\end{eqnarray}
where $t_{ij}$ are matrix elements of $T$. From
Eq.(\ref{2mn-mat-pair}) and Eq.(\ref{2mn-mat-pair-ILO}) we can see
that the SLOCC equivalence of the quantum state turns to the
connectivity of the matrix pairs $(\Gamma_{\!1}, \Gamma_{\!2})$
under the special linear transformations $T,P,Q$. Define the set
that contains all the matrices pair $(\Gamma_{\!1},\Gamma_{\!2})$ as
$C$. The whole space of $C$ can be partitioned into numbers of
subsets with different $n$, $l$
\begin{eqnarray}
C_{n,\;l}=\{ (\Gamma_{\!1},\Gamma_{\!2})|\;
r_{\mathrm{max}}(\alpha_{1} \Gamma_{\!1} + \beta_{1} \Gamma_{\!2}) =
n, r_{\mathrm{min}}(\alpha_{2} \Gamma_{\!1} + \beta_{2}
\Gamma_{\!2}) = l \} \; , \label{nldef}
\end{eqnarray}
where $r_{\rm max}$ and $r_{\rm min}$ represent the the maximum and
minimum rank of the matrices respectively; $\alpha_{i}, \beta_{i}
\in \mathbb{C} $ and $|\alpha_{i}| + |\beta_{i}| \neq 0$; $l\in
[0,n]$, $n \in [0, M]$.
\begin{proposition}
If $(\Gamma_1,\Gamma_2) \in C_{n,l}$ and $\exists$ $T,P,Q \in \mathrm{ILO}$,
$\begin{pmatrix}\Gamma'_1 \\
\Gamma'_2
\end{pmatrix} = \begin{pmatrix}
    t_{11} & t_{12} \\
    t_{21} & t_{22} \\
\end{pmatrix} \begin{pmatrix}P\Gamma_1Q \\ P\Gamma_2Q
\end{pmatrix} $,
then $(\Gamma'_1,\Gamma'_2) \in C_{n,l}$. \label{proposition1}
\end{proposition}
(see Appendix \ref{appdixproposition1}). This proposition implies
that the matrix pairs in subsets $C_{n,l}$ with different $n$ or $l$
are SLOCC inequivalent.

\subsection{Classification on sets $C_{n,l}$ with $n=M$} \label{cml}

We start our classification of $C_{n,\,l}$ in $2\times M\times N$
system from the case $n=M$. Our aim is to construct the subsets
$c_{M,l}\subset C_{n,\,l}$ which: (i), it includes representative
states of all the inequivalent entanglement classes; (ii), each
inequivalent class has only one representative state in $c_{M,l}$.

Because $\forall$ $(\Gamma_{\!1},\Gamma_{\!2})\in C_{M,\, l}$,
$\exists$ $T\in \mathrm{ILO}$ (see Appendix
\ref{appdixproposition1})
\begin{eqnarray}
T \left\lgroup
\begin{array}{c}
\Gamma_{\!1} \\
\Gamma_{\!2} \\
\end{array}
\right\rgroup =
\left\lgroup
\begin{array}{cc}
    t_{11} & t_{12} \\
    t_{21} & t_{22} \\
\end{array}
\right\rgroup
\left\lgroup
\begin{array}{c}
    \Gamma_{\!1} \\
    \Gamma_{\!2} \\
\end{array}
\right\rgroup  , \label{transf}
\end{eqnarray}
that makes $ r(t_{11} \Gamma_{\!1} + t_{12} \Gamma_{\!2}) = M$, $
r(t_{21}\Gamma_{\!1} + t_{22}\Gamma_{\!2}) = l$, so we assume that
all the matrix pairs in $C_{M,l}$ have been performed this kind of
ILO transformation $T$. That is $r(\Gamma_{\!1}) = M$ and
$r(\Gamma_{\!2})=l$. Two specific ILOs $P$ and $Q$ can transform
$(\Gamma_{\!1},\Gamma_{\!2})$ into the following form
\begin{eqnarray}
\left( \begin{array}{l} \Gamma_{\!1} \\ \Gamma_{\!2}
\end{array} \right) \xrightarrow{P,Q} \left(
\begin{array}{ll}
\left( \begin{array}{ll} E_{\stl M\times M}
& \mathbf{0}_{\stl M\times (N-M)} \end{array}\right) \\
\left(\begin{array}{ll} A_{\stl M\times M} & B_{\stl M\times (N-M)}
\end{array} \right)
\end{array}
\right)\; ,\label{EOAB}
\end{eqnarray}
where $E$ is an unit submatrix of $P\Gamma_{\!1}Q$, $\mathbf{0}$ is
zero submatrix; $A$ and $B$ are submatrix of $P\Gamma_2Q$, and all
of them have the subscripts as their dimensions. We can represent
the submatrix $B_{M\times (N-M)}$ by matrix theory conventions,
i.e., $B_{M\times (N-M)} = \Gamma_{2}(\{1,\cdots, M\},\{ M+1,
\cdots, N \})$.

If $(N-M)>M$, then $r_{\text{max}}(B_{M\times(N-M)})=M$, the right
hand of Eq.(\ref{EOAB}) can be further transformed by ILOs into
\begin{eqnarray}
\left(\begin{array}{c} \Gamma_{\!1} \\ \Gamma_{\!2}
\end{array} \right) \xrightarrow{P,Q} \left(
\begin{array}{l}
\left( \begin{array}{lll} E_{\stl M\times M} & \mathbf{0}_{\stl
M\times (N-2M)} &
\mathbf{0}_{\stl M\times M} \end{array}\right) \\
\left( \begin{array}{lll} \mathbf{0}_{\stl M\times M} &
\mathbf{0}_{\stl M\times (N-2M)} & E_{\stl M\times M}
\end{array}\right)
\end{array}
\right)\; , \label{empty-planes}
\end{eqnarray}
In the form of the cubic grid (Fig.(\ref{pic-cube-234})), this
corresponds to that at least $(N-2M)$ vertical planes in the middle
of the cube are zero planes, which is actually an entangled states
of $2\times M\times 2M$ according to lemma \ref{lemma1}. Thus here
we consider the case $M \geq N/2$.

For arbitrary matrix pair with the form of the right hand of
Eq.(\ref{EOAB}), we implement the following transformation via ILOs
\begin{eqnarray}
\left(
\begin{array}{ll}
\left( \begin{array}{ll} E_{\stl m\times m}
& \mathbf{0}_{\stl m\times (n-m)} \end{array}\right) \\
\left(\begin{array}{ll} A_{\stl m\times m} & B_{\stl m\times (n-m)}
\end{array} \right)
\end{array}
\right) \xrightarrow{ \text{step {\bf i}}}  \left(
\begin{array}{l} \left(\begin{array}{lll}
E_{\stl 1A'} & \mathbf{0}_{\stl 1B'} & \mathbf{0}_{\stl 1a}  \\
\mathbf{0}_{\stl 1b} & E_{\stl 1'} & \mathbf{0}_{\stl 1E'}
\end{array}
\right) \\ \left(
\begin{array}{lll}
A' & B' & \mathbf{0}_{\stl 2a}  \\
\mathbf{0}_{\stl 2b} & \mathbf{0}_{\stl 2c} & E'
\end{array}
\right)\end{array} \right) \; , \label{gam1-form1}
\end{eqnarray}
where the lower-right submatrix of the right hand side
$\Gamma_{\!2}(\{m-r(B)+1,\cdots,m\},\{m+1,\cdots, n\})= E'$ has
$r(E')=r(B)$; $A'$, $E_{\stl 1A'}$ are square submatrices with the
dimensions $(m-r(E')) \times (m-r(E'))$; the rest of the matrices
are partitioned accordingly, i.e., $\mathbf{0}_{\stl 1B'}$, $B'$
have the dimension $(m-r(E')) \times r(E')$, $\mathbf{0}_{1a}$,
$\mathbf{0}_{2a}$ have the dimension of $(m-r(E'))\times (n-m)$,
$\mathbf{0}_{\stl 1b}$, $\mathbf{0}_{2b}$ have the dimension $r(E')
\times (m-r(E'))$, $E_{\stl 1'}$, $\mathbf{0}_{\stl 2c}$ have the
dimension $r(E')\times r(E')$. After the transformation,
$\Gamma_{\!1} = (E_{\stl M\times M}, \mbd{0}_{\stl M\times (N-M)})$
being unchanged, $\Gamma_{\!2}$ becomes a quasidiagonal matrix and
we named this procedure step {\bf i}.

Next we repartitioned the matrices on the left hand side of
Eq.(\ref{gam1-form1}) as follows
\begin{eqnarray}
\left(
\begin{array}{l} \left(\begin{array}{lll}
E_{\stl 1A'} & \mathbf{0}_{\stl 1B'} & \mathbf{0}_{\stl 1a}  \\
\mathbf{0}_{\stl 1b} & E_{\stl 1'} & \mathbf{0}_{\stl 1E'}
\end{array}
\right) \\ \left(
\begin{array}{lll}
A' & B' & \mathbf{0}_{\stl 2a}  \\
\mathbf{0}_{\stl 2b} & \mathbf{0}_{\stl 2c} & E'
\end{array}
\right) \end{array} \right) \xrightarrow{\text{step {\bf ii}}}
\left(
\begin{array}{l} \left(\begin{array}{ll|l}
E_{\stl 1A'} & \mathbf{0}_{\stl 1a} & \mathbf{0}_{\stl 1b}  \\
\hline \mathbf{0}_{\stl 1c} & E_{\stl 1'} & \mathbf{0}_{\stl 1d}
\end{array}
\right) \\ \left(
\begin{array}{ll|l}
A' & B' & \mathbf{0}_{\stl 2a}  \\ \hline \mathbf{0}_{\stl 2b} &
\mathbf{0}_{\stl 2c} & E'
\end{array}
\right)\end{array} \right) \; . \label{partition-AB}
\end{eqnarray}
This is named as step {\bf ii}. Consider the submatrix $B'$, if it
is not identically zero we can perform the transformation of step
{\bf i} on the left-top submatrices $\begin{pmatrix}A' & B'
\end{pmatrix}$ of Eq.(\ref{partition-AB})
\begin{eqnarray}
\left(
\begin{array}{ll}
\left( \begin{array}{ll} E_{\stl 1A'} & \mathbf{0}_{\stl 1a} \end{array}\right) \\
\left(\begin{array}{ll} A' & B'
\end{array} \right)
\end{array}
\right) \xrightarrow{\text{step {\bf i}}} \left(
\begin{array}{l} \left(\begin{array}{lll}
E_{\stl 1A''} & \mathbf{0}_{\stl 1B''} & \mathbf{0}_{\stl 1a'}  \\
\mathbf{0}_{\stl 1b'} & E_{\stl 1''} & \mathbf{0}_{\stl 1E''}
\end{array}
\right) \\ \left(
\begin{array}{lll}
A'' & B'' & \mathbf{0}_{\stl 2a}  \\
\mathbf{0}_{\stl 2b'} & \mathbf{0}_{\stl 2c'} & E''
\end{array}
\right)\end{array} \right) \; .
\end{eqnarray}
This procedure can be done repeatedly (suppose repeat $n$ times),
until the $r(B^{(n)})=0$. We can get that the matrix pair
$(\Gamma_{\!1},\Gamma_{\!2})$ can be transformed into the following
form
\begin{eqnarray}
\Gamma_{\!1} \rightarrow \left(
\begin{array}{l|lllll}
E_{\stl 1A^{(n)}} & \mathbf{0} & \mathbf{0} & \cdots & \mathbf{0} &
\mathbf{0} \\ \hline \mathbf{0} & E_{ 1^{
(n-1)}} & \mathbf{0} & \cdots & \mathbf{0} & \mathbf{0} \\
\vdots & \vdots & \vdots & \ddots & \vdots & \vdots \\ \mathbf{0} &
\mathbf{0} &
\mathbf{0} & \cdots & \mathbf{0} & \mathbf{0} \\
\mathbf{0} & \mathbf{0} & \mathbf{0} & \cdots & E_{ 1'} & \mathbf{0}
\end{array}
\right) & \equiv & \left(
\begin{array}{ll}
E & \mathbf{0}  \\
\mathbf{0} & E_1
\end{array}\right) \; , \label{normal-G1}
\end{eqnarray}
\begin{eqnarray}
\Gamma_{\!2} \rightarrow  \left(
\begin{array}{l|lllll}
A^{\stl (n)} & B^{\stl (n)} = \mathbf{0} & \mathbf{0}
& \cdots & \mathbf{0} & \mathbf{0} \\
\hline \mathbf{0} & \mathbf{0} & E^{\stl (n-1)}
& \cdots & \mathbf{0} & \mathbf{0} \\
\vdots & \vdots & \vdots & \ddots & \vdots & \vdots \\ \mathbf{0} &
\mathbf{0}
& \mathbf{0} & \cdots & E'' & \mathbf{0} \\
\mathbf{0} & \mathbf{0} & \mathbf{0} & \cdots & \mathbf{0} & E'
\end{array}
\right) \equiv \left(\begin{array}{ll}
SJS^{-1} & \mathbf{0}  \\
\mathbf{0} & E_2
\end{array}\right) \; , \label{normal-G2}
\end{eqnarray}
where the transformed $\Gamma_{\!1}$ is just $(E_{\stl M\times M},
\mbd{0}_{M\times (N-M)})$, and $E_{1,2}$ are lower-right submatrices
defined according to the partition lines; $J$ is the Jordan form of
$A^{(n)}$.

As a concrete example here we show how this whole procedure is
proceeded on the sets of $C_{4,l}$ of $2\times 4\times 6$ state. The
transformation of Eq.(\ref{gam1-form1}) is start with
\begin{eqnarray}
\left(
\begin{array}{ll}
\left( \begin{array}{ll} E_{\stl 4\times 4} & \mathbf{0}_{\stl 4\times 2} \end{array}\right) \\
\left(\begin{array}{ll} A_{\stl 4\times 4} & B_{\stl 4\times 2}
\end{array} \right)
\end{array}
\right) \xrightarrow{\text{step {\bf i}}}  \left(
\begin{array}{l} \left(\begin{array}{lll}
E_{\stl 1A'} & \mathbf{0}_{\stl 1B'} & \mathbf{0}_{\stl 1a}  \\
\mathbf{0}_{\stl 1b} & E_{\stl 1'} & \mathbf{0}_{\stl 1c}
\end{array}
\right) \\ \left(
\begin{array}{lll}
A' & B' & \mathbf{0}_{\stl 2a}  \\
\mathbf{0}_{\stl 2b} & \mathbf{0}_{\stl 2c} & E'
\end{array}
\right)\end{array} \right) \; ,
\end{eqnarray}
where
\begin{eqnarray}
\left(
\begin{array}{l} \left(\begin{array}{lll}
E_{\stl 1A'} & \mathbf{0}_{\stl 1B'} & \mathbf{0}_{\stl 1a}  \\
\mathbf{0}_{\stl 1b} & E_{\stl 1'} & \mathbf{0}_{\stl 1c}
\end{array}
\right) \\ \left(
\begin{array}{lll}
A' & B' & \mathbf{0}_{\stl 2a}  \\
\mathbf{0}_{\stl 2b} & \mathbf{0}_{\stl 2c} & E'
\end{array}
\right)\end{array} \right) =  \left(
\begin{array}{l} \left(\begin{array}{ll|ll|ll}
1 & 0 & 0 & 0 & 0 & 0 \\
0 & 1 & 0 & 0 & 0 & 0 \\ \hline
0 & 0 & 1 & 0 & 0 & 0 \\
0 & 0 & 0 & 1 & 0 & 0
\end{array}
\right) \\ \left(\begin{array}{ll|ll|ll}
\times & \times & + & + & 0 & 0 \\
\times & \times & + & + & 0 & 0 \\ \hline
0 & 0 & 0 & 0 & 1 & 0 \\
0 & 0 & 0 & 0 & 0 & 1
\end{array}
\right) \end{array} \right) \; .
\end{eqnarray}
Here, the rank of $B_{\stl 4\times 2}$ must be 2, otherwise the
state will not be a true entangled state of $2\times 4\times 6$,
similar to the argument below Eq.(\ref{empty-planes}). The step {\bf
ii} goes as follows
\begin{eqnarray}
\left(
\begin{array}{l} \left(\begin{array}{lll}
E_{\stl 1A'} & \mathbf{0}_{\stl 1B'} & \mathbf{0}_{\stl 1a}  \\
\mathbf{0}_{\stl 1b} & E_{\stl 1'} & \mathbf{0}_{\stl 1c}
\end{array}
\right) \\ \left(
\begin{array}{lll}
A' & B' & \mathbf{0}_{\stl 2a}  \\
\mathbf{0}_{\stl 2b} & \mathbf{0}_{\stl 2c} & E'
\end{array}
\right)\end{array} \right) \xrightarrow{\text{step ii}} \left(
\begin{array}{l} \left(\begin{array}{ll|l}
E_{\stl 1A'} & \mathbf{0}_{\stl 1B'} & \mathbf{0}_{\stl 1a}
\\ \hline \mathbf{0}_{\stl 1b} & E_{\stl 1'} & \mathbf{0}_{\stl 1c}
\end{array}
\right) \\ \left(
\begin{array}{ll|l}
A' & B' & \mathbf{0}_{\stl 2a}  \\ \hline \mathbf{0}_{\stl 2b} &
\mathbf{0}_{\stl 2c} & E'
\end{array}
\right)\end{array} \right) \; . \label{eg246ii}
\end{eqnarray}

Next we repeat the step {\bf i} to the up-left submatrices of the
right hand side of Eq.(\ref{eg246ii}). This iteration of step {\bf
i} depends on the rank of $B'$.

\noindent (1), $r(B') = 0$. In this case the matrix pair
$(\Gamma_{\!1},\Gamma_{\!2})$ become
\begin{eqnarray}
\left( \begin{array}{l} \Gamma_{\!1} \\ \Gamma_{\!2}
\end{array} \right) \xrightarrow{T,P,Q}
\left(
\begin{array}{l} \left(\begin{array}{ll|ll|ll}
1 & 0 & 0 & 0 & 0 & 0 \\
0 & 1 & 0 & 0 & 0 & 0 \\ \hline
0 & 0 & 1 & 0 & 0 & 0 \\
0 & 0 & 0 & 1 & 0 & 0
\end{array}
\right) \\ \left(\begin{array}{ll|ll|ll}
\times & \times & 0 & 0 & 0 & 0 \\
\times & \times & 0 & 0 & 0 & 0 \\ \hline
0 & 0 & 0 & 0 & 1 & 0 \\
0 & 0 & 0 & 0 & 0 & 1
\end{array}
\right) \end{array} \right) \; .
\end{eqnarray}
And there are three different forms of $\Gamma_2$, i.e.,
\begin{eqnarray}
(1.1) \left[
\begin{array}{llllll}
0 & 0 & 0 & 0 & 0 & 0 \\
0 & 0 & 0 & 0 & 0 & 0 \\
0 & 0 & 0 & 0 & 1 & 0 \\
0 & 0 & 0 & 0 & 0 & 1
\end{array}
\right],\; (1.2)  \left[
\begin{array}{llllll}
\lambda & 0 & 0 & 0 & 0 & 0 \\
0 & 0 & 0 & 0 & 0 & 0 \\
0 & 0 & 0 & 0 & 1 & 0 \\
0 & 0 & 0 & 0 & 0 & 1
\end{array}
\right],\; (1.3) \left[
\begin{array}{llllll}
0 & 1 & 0 & 0 & 0 & 0 \\
0 & 0 & 0 & 0 & 0 & 0 \\
0 & 0 & 0 & 0 & 1 & 0 \\
0 & 0 & 0 & 0 & 0 & 1
\end{array}
\right] , \nonumber
\end{eqnarray}
correspond to two Jordan canonical forms of $A'$, $J = \left[
\begin{array}{ll}
\lambda & 0 \\
0 & 0
\end{array}
\right]$, $J = \left[
\begin{array}{ll}
0 & 1 \\
0 & 0
\end{array}
\right]$, and a zero matrix $A' = \left[
\begin{array}{ll}
0 & 0 \\
0 & 0
\end{array}
\right]$.

\noindent (2), $r(B') = 1$. In this case
\begin{eqnarray}
\left(
\begin{array}{lll}
A' & B' & \mathbf{0} \\  \mathbf{0} & \mathbf{0} & E'
\end{array}
\right)  \xrightarrow{\text{ step {\bf i}}} \left(
\begin{array}{ll|ll|l}
\times & \times & 0 & 0 & \mathbf{0}  \\
0 & 0 & 0 & 1 & \mathbf{0}  \\ \hline \mathbf{0} & \mathbf{0} &
\mathbf{0} & \mathbf{0} & E'
\end{array}
\right)\; . \label{rb=1}
\end{eqnarray}

\begin{eqnarray}
\left(
\begin{array}{ll|ll|l}
\times & \times & 0 & 0 & \mbd{0}  \\  0 & 0 & 0 & 1 & \mathbf{0}
\\ \hline \mathbf{0} & \mathbf{0} & \mathbf{0} & \mathbf{0} & E'
\end{array}
\right) \xrightarrow{\text{step {\bf ii}}} \left(
\begin{array}{ll|ll}
A'' & B'' & \mbd{0} & \mathbf{0}  \\ \hline 0 & 0 & E'' & \mathbf{0}
\\  \mathbf{0} & \mathbf{0} & \mathbf{0} & E'
\end{array}
\right) \; ,
\end{eqnarray}
where $A''$, $B''$ are matrices of $1\times 1$ and $E''=(0,1)$.
Again apply step {\bf i} on $(A''\; B'')$ we have

\noindent (2.1), $r(B'') = 0$
\begin{eqnarray}
\left(
\begin{array}{ll|ll}
A'' & B'' & \mbd{0} & \mbd{0}  \\ \hline 0 & 0 & E'' & \mathbf{0}
\\  \mathbf{0} & \mathbf{0} & \mathbf{0} & E'
\end{array}
\right)  & \xrightarrow{\text{ step {\bf i}}} & \left(
\begin{array}{ll|ll}
\times & 0 & \mbd{0} & \mathbf{0}  \\ \hline 0 & 0  & E'' &
\mathbf{0}
\\  \mathbf{0} & \mathbf{0} & \mathbf{0} & E'
\end{array}
\right) \; . \label{bpp=0}
\end{eqnarray}

\noindent (2.2)~$r(B'') = 1$
\begin{eqnarray}
\left(
\begin{array}{ll|ll}
A'' & B'' & \mbd{0} & \mbd{0}  \\ \hline 0 & 0 & E'' & \mathbf{0}
\\  \mathbf{0} & \mathbf{0} & \mathbf{0} & E'
\end{array}
\right)  & \xrightarrow{\text{ step {\bf i}}} & \left(
\begin{array}{ll|ll}
0 & 1  & \mbd{0} & \mathbf{0}  \\ \hline 0 & 0 & E'' & \mathbf{0}  \\
\mathbf{0} & \mathbf{0}  & \mathbf{0} & E'
\end{array}
\right) \; . \label{bpp=1}
\end{eqnarray}
For Eq.(\ref{bpp=0}), $A''$ is equivalent to the case of $A''=0$
according to {\it theorem 1} of \cite{class-nn}. For
Eq.(\ref{bpp=1}), in the next step of  step {\bf ii}, $B^{(3)}$ will
be a matrix of dimension zero, and satisfies $r(B^{(3)})=0$, thus
the procedure is stopped. We get two inequivalent forms of
$\Gamma_{\!2}$
\begin{eqnarray}
\left[
\begin{array}{llllll}
0 & 0 & 0 & 0 & 0 & 0 \\
0 & 0 & 0 & 1 & 0 & 0 \\
0 & 0 & 0 & 0 & 1 & 0 \\
0 & 0 & 0 & 0 & 0 & 1
\end{array}
\right], \left[
\begin{array}{llllll}
0 & 1 & 0 & 0 & 0 & 0 \\
0 & 0 & 0 & 1 & 0 & 0 \\
0 & 0 & 0 & 0 & 1 & 0 \\
0 & 0 & 0 & 0 & 0 & 1
\end{array}
\right]\; .
\end{eqnarray}

\noindent (3). $r(B') = 2$. In this case
\begin{eqnarray}
\left(
\begin{array}{lll}
A' & B' & \mathbf{0} \\  \mathbf{0} & \mathbf{0} & E'
\end{array}
\right) \xrightarrow{\text{step {\bf i}}}  \left(
\begin{array}{ll|ll|l}
0 & 0 & 1 & 0 & \mathbf{0}  \\
0 & 0 & 0 & 1 & \mathbf{0}  \\ \hline \mathbf{0} & \mathbf{0} &
\mathbf{0} & \mathbf{0} & E'
\end{array}
\right) \; . \label{3gamma1}
\end{eqnarray}
Thus here is only one class, where $\Gamma_{\!2}$ has just the form
of Eq.(\ref{3gamma1}). In the following, we shall see that these six
cases correspond to the six inequivalent entanglement classes in
$2\times 4\times 6$ systems, which agrees with the result of
Ref.\cite{range-2mn}.

In all, for every $(\Gamma_{\!1},\Gamma_{\!2})\in C_{M,\; l}$, there
exists an ILO transformation that make
\begin{eqnarray}
\begin{pmatrix}
    \Gamma_{\!1}' \\
    \Gamma_{\!2}' \end{pmatrix} = T \otimes P \otimes Q
\begin{pmatrix}
\Gamma_{\!1} \\
\Gamma_{\!2} \end{pmatrix} \; . \label{N-transf}
\end{eqnarray}
Here $\Gamma_{\!1}'$ has the form of Eq.(\ref{normal-G1}), and
$\Gamma_{\!2}' = \left(\begin{array}{ll} J & 0 \\ 0 & E_2
\end{array}\right)$ has the form of Eq.(\ref{normal-G2}). Eq.(\ref{N-transf}) maps
$C_{M,\,l}$ to $c_{M,\,l}$, where $c_{M,\;l} \subseteq C_{M,\;l}$
and
\begin{eqnarray}
c_{M,\,l} = \{ (\Gamma_{\!1},\Gamma_{\!2})|\Gamma_{\!1}=\left(
\begin{array}{ll}
E & \mathbf{0}  \\
\mathbf{0} & E_1
\end{array}\right) ,
\Gamma_{\!2}= \left(\begin{array}{ll}
J & \mathbf{0} \\
\mathbf{0} & E_2
\end{array}\right); (\Gamma_{\!1},\Gamma_{\!2}) \in C_{M,\,l} \}\; .
\label{set-CN}
\end{eqnarray}
Thus we have separated the classification of $C_{M,l}$ into two
procedures: (1), the construction of $E_2$ matrix; (2),
classification of $J$. And for the second procedure, we have already
completed the classification in \cite{class-nn}. We have the
following theorem

\begin{theorem}
$\forall$ $(\Gamma_{\!1},\Gamma_{\!2}) \in c_{M,\;l}$, the set
$c_{M,\; l}$ is of the classification of $C_{M,\;l}$. $\mathrm{(i)}$
if two states are SLOCC equivalent then they can be transformed into
the same matrix vector $(\Gamma_{\!1},\Gamma_{\!2})$;
$\mathrm{(ii)}$ this matrix vector is unique in $c_{M,\;l}$, that is
if $(\Gamma_{\!1},\Gamma_{\!2}')$ is SLOCC equivalent with
$(\Gamma_{\!1},\Gamma_{\!2})$, then $(\Gamma_{\!1},\Gamma_{\!2}') =
(\Gamma_{\!1},\Gamma_{\!2})$, $(\Gamma_{\!2}'=\Gamma_{\!2}$ means
that $E_2=E_2'$ and their Jordan forms of $J$ are equivalent under
the condition of theorem 1 Ref.\cite{class-nn} $)$
\end{theorem}

\noindent Proof:\\ \\ (i) The proof of this statement is
straightforward, since in every step of transformation only
invertible operators take part in.

\noindent
(ii) Suppose
\begin{eqnarray}
\left(\begin{array}{l}
\Gamma_{\!1} \\
\Gamma_{\!2}'
\end{array}\right) = T'\otimes P'\otimes Q'
\left(
\begin{array}{l}
\Gamma_{\!1} \\
\Gamma_{\!2}
\end{array}
\right) \; . \label{tpq}
\end{eqnarray}
It can be proved that the $T'$ transformations can always be
replaced by ILO operators $P_{0}^{-1},Q_{0}^{-1}$, i.e., (see
Appendix \ref{TtoPQ})
\begin{eqnarray}
\left\lgroup
  \begin{array}{cc}
    t'_{11} & t'_{12} \\
    t'_{21} & t'_{22} \\
  \end{array}
\right\rgroup
\left(
\begin{array}{l}
\Gamma_{\!1} \\
\Gamma_{\!2}
\end{array}
\right) = \left(
\begin{array}{l}
P_{0}^{-1} \Gamma_{\!1}Q_{0}^{-1} \\
P_{0}^{-1} \Gamma_{\!2} Q_{0}^{-1}
\end{array}
\right) \label{t-trans}\; .
\end{eqnarray}
Thus Eq.(\ref{tpq}) can be rewritten as
\begin{eqnarray}
\left(\begin{array}{l}
\Gamma_{\!1} \\
\Gamma_{\!2}'
\end{array}\right) & = & P'P_{0}^{-1} \left(\begin{array}{l}
\Gamma_{\!1} \\
\Gamma_{\!2}
\end{array} \right) Q_{0}^{-1}Q' \nonumber \\ & = & P'' \left(\begin{array}{l}
\Gamma_{\!1} \\
\Gamma_{\!2}
\end{array}\right)Q'' \; ,
\end{eqnarray}
which correspond to two matrix equations
\begin{eqnarray}
\left\{ \begin{array}{l} P''\Gamma_1 Q'' = \Gamma_1 \\ \\
P''\Gamma_2 Q'' = \Gamma_2'
\end{array} \right. \; . \label{matr-equa1}
\end{eqnarray}
We proceed our proof along the procedure of the construction of the
standard form of $c_{M,l}$. When $\Gamma_{\!1}$ has the form of the
left hand side of Eq.(\ref{gam1-form1}), the invertible
transformation $P'', Q''$ that keep it invariant must be of the form
\begin{eqnarray}
Q''  = \left(
\begin{array}{ll}
P''^{-1} & \mathbf{0} \\
X & Y
\end{array} \right) \; , \label{B-inden1}
\end{eqnarray}
where $\mathrm{Det}(Y)\neq 0$. This transformation transform
$\Gamma_{\!2}$ of the left hand side of Eq.(\ref{gam1-form1}) into
the follow
\begin{eqnarray}
P''\Gamma_{\!2}Q'' = P''(A,B)Q'' = (P''AP''^{-1}+P''B X,P''B Y)\; ,
\label{argu-equiv}
\end{eqnarray}
where $A$ is the $ M\times M$ submatrix, and $B$ is the $ M \times
(N-M)$ submatrix. Since $P''$ and $Y$ both are ILO operators, the
rank of submatrix $B$, is unchanged and it can be further
transformed to form of the right hand side of Eq.(\ref{gam1-form1})
\begin{eqnarray}
\left(\begin{array}{lll}
A' & B' & \mathbf{0} \\
\mathbf{0} & \mathbf{0} & E'
\end{array}\right)\; . \label{decup}
\end{eqnarray}
We get that if two states are SLOCC equivalent then $E'$ block of
$\Gamma_{\!2}'$ and $\Gamma_{\!2}$ must be identical. In
Eq.(\ref{decup}) we see that Eq.(\ref{decup}) can be partitioned as
the step {\bf ii} in Eq.(\ref{partition-AB}). Then we apply the same
argument as Eqs.(\ref{B-inden1},\ref{argu-equiv}) on submatrix
$\begin{pmatrix}A' & B'\end{pmatrix}$. We can arrive that the $E''$
( $E^{(3)}$, $E^{(4)}$ and so on) must also be identical according
to Eq.(\ref{matr-equa1}). And finally we can get that if
$(\Gamma_{\!1},\Gamma_{\!2}')$ is SLOCC equivalent with
$(\Gamma_{\!1},\Gamma_{\!2})$ then $\Gamma_{\!2}'$ and
$\Gamma_{\!2}$ have the same canonical form in the set of
Eq.(\ref{set-CN}). Q.E.D.

\subsection{Classification on sets $C_{n,l}$ with $n=M-i$}

Here we start by constructing the standard form of the set
$C_{M-i,l}$ using ILOs. It is shown that the construction of the
entanglement classes $c_{M-i,l}$ can be realized by apply the
transformations of $c_{M,l}$ on both columns and rows of the matrix
pairs $(\Gamma_1, \Gamma_2)$.

$\forall (\Gamma_1,\Gamma_2) \in C_{M-i,l}$, $(\Gamma_1,\Gamma_2)$
can be transformed into the following form
\begin{eqnarray}
\begin{pmatrix}\Gamma_1 \\ \Gamma_2\end{pmatrix} \xrightarrow{T,P,Q}
\begin{pmatrix}\Gamma_1 \\ \Gamma_2\end{pmatrix}=
\begin{pmatrix}
\left(\begin{array}{l|ll} E_{(M-i)\times (M-i)} & \mathbf{0} &
\mathbf{0} \\ \hline \mathbf{0} & \mathbf{0}_{i\times i} &
\mathbf{0}_{i\times (N-M)}
\end{array}\right) \\
\left(\begin{array}{l|ll} \times_{(M-i)\times (M-i)} & \times &
\times \\ \hline \times & \mathbf{0}_{i\times i} &
\mathbf{0}_{i\times (N-M)}
\end{array}\right)
\end{pmatrix} \; , \label{Gamma1-S-Form}
\end{eqnarray}
where $\Gamma_2$ is partitioned according to the partitions of
$\Gamma_1$. Here due to $r_{\text{max}}(\alpha_1\Gamma_1 +
\beta_1\Gamma_2) = M-i$, submatrix
$\Gamma_2(\{M-i+1,\cdots,M\},\{M-i+1,\cdots,N\})$ must be zero
matrix. After this transformation, we can apply the step {\bf i} in
Eq.(\ref{gam1-form1}) on the submatrices $\Gamma_2(\{1,
\cdots,M-i\},\{ 1,\cdots, N \})$ and $\Gamma_2(\{1, \cdots, M\},\{
1,\cdots, M-i \})$ on the right hand side of
Eq.(\ref{Gamma1-S-Form}). $\Gamma_2$ then turns to  (see Appendix
{\ref{App-C}})
\begin{eqnarray}
\Gamma_{\!2}\xrightarrow{\text{ILO}}  \left(
\begin{array}{llllll|ll}
\times & \times & \times & \mbd{0} & \times & \times & \mbd{0} & \mbd{0} \\
\times & \times & \times & \mbd{0} & \times & \times & \mbd{0} & \mbd{0} \\
\times & \times & \times & \mbd{0} & \times & \times & \mbd{0} & \mbd{0} \\
\times & \times & \times & \mbd{0} & \mbd{0}_{i\times i} & \mbd{0} &
\mbd{0}
& \mbd{0} \\
\mbd{0} & \mbd{0} & \mbd{0} & \mbd{0} & \mbd{0} & \mbd{0} & E_{i\times i} & \mbd{0} \\
\mbd{0} & \mbd{0} & \mbd{0} & \mbd{0} & \mbd{0} & \mbd{0} & \mbd{0} & E_{(N-M)\times(N-M)} \\
\hline \mbd{0} & \mbd{0} & \mbd{0} & E_{i\times i} & \mbd{0} &
\mbd{0} & \mbd{0}_{i\times i} & \mbd{0}_{i\times (N-M)}
\end{array}
\right), \label{cm-iGamma-2}
\end{eqnarray}
while $\Gamma_1$ being unchanged. Repartition the above equation as
follows
\begin{eqnarray}
\left(
\begin{array}{lll|lllll}
\times & \times & \times & \mbd{0} & \times & \times & \mbd{0} & \mbd{0} \\
\times & \times & \times & \mbd{0} & \times & \times & \mbd{0} & \mbd{0} \\
\times & \times & \times & \mbd{0} & \times & \times & \mbd{0} & \mbd{0} \\
\hline
\times & \times & \times & 0 & 0_{i\times i} & \mbd{0} & \mbd{0} & \mbd{0} \\
\mbd{0} & \mbd{0} & \mbd{0} & \mbd{0} & \mbd{0} & \mbd{0} & E_{i\times i} & \mbd{0} \\
\mbd{0} & \mbd{0} & \mbd{0} & \mbd{0} & \mbd{0} &
\mbd{0}_{(N-M)\times(N-M)} & \mbd{0} & E_{(N-M)\times(N-M)}
\\  \mbd{0} & \mbd{0} & \mbd{0} & E_{i\times i} & \mbd{0} & \mbd{0} & \mbd{0}_{i\times i} & \mbd{0}_{i\times
(N-M)}
\end{array}
\right),
\end{eqnarray}
where the lower-right submatrix is $(3i+N-M)\times (3i+2(N-M))$.
\begin{proposition}
There exists true entanglement state in $2\times M\times N$ pure
systems if and only if $(\Gamma_1,\Gamma_2) \in C_{n,l}$ where $n
\geq \frac{M+N}{3}$. \label{proposition2}
\end{proposition}
This proposition reduce to the Eq.(81) of \cite{class-nn} when
$M=N$. Let $ \eta = \{ 1,\cdots, 2M-2i-N \} $, $\rho =
\{1,\cdots,2M-3i-N, 2M-2i-N+1,\cdots, M-i\}$, then
\begin{eqnarray}
(\Gamma_1,\Gamma_2)( \eta , \rho ) = \left(
\begin{array}{lll|ll}
\times & \times & \times & \times & \times  \\
\times & \times & \times & \times & \times  \\
\times & \times & \times & \times & \times  \\ \hline \times &
\times & \times &  \mbd{0}_{i\times i} & \mbd{0}_{i\times (N-M)}
\end{array}
\right),
\end{eqnarray}
has the same structure as the right hand side of
Eq.(\ref{Gamma1-S-Form}), where $\Gamma(\eta,\rho)$ is the submatrix
of $\Gamma$ with the selected rows and columns in sets $\eta$ and
$\rho$, separately. Then we can apply the same procedure as that of
Eq.(\ref{Gamma1-S-Form}).

Here presents the $2\times 7\times 8$ state as a demonstration,
i.e., $C_{ M-1,\,l} = C_{\stl 6,l}$. The matrix pair
$(\Gamma_{\!1},\Gamma_{\!2})$ can be transformed into the following
form
\begin{eqnarray}
\left(\begin{array}{l} \Gamma_{\!1} \\ \Gamma_{\!2}
\end{array}\right)  \xrightarrow{T,\, P,\, Q} \left(\begin{array}{l} \left(
\begin{array}{llllllll}
1 & 0 & 0 & 0 & 0 & 0 & 0 & 0 \\
0 & 1 & 0 & 0 & 0 & 0 & 0 & 0 \\
0 & 0 & 1 & 0 & 0 & 0 & 0 & 0 \\
0 & 0 & 0 & 1 & 0 & 0 & 0 & 0 \\
0 & 0 & 0 & 0 & 1 & 0 & 0 & 0 \\
0 & 0 & 0 & 0 & 0 & 1 & 0 & 0 \\
0 & 0 & 0 & 0 & 0 & 0 & 0 & 0
\end{array}
\right) \\ \left(
\begin{array}{lll|lllll}
\times & \times & \times & 0 & c_{04} & c_{05} & 0 & 0 \\
\times & \times & \times & 0 & c_{14} & c_{15} & 0 & 0 \\
\times & \times & \times & 0 & c_{24} & c_{25} & 0 & 0 \\ \hline
r_{30} & r_{31} & r_{32} & 0 & 0 & 0 & 0 & 0 \\
0 & 0 & 0 & 0 & 0 & 0 & 1 & 0 \\
0 & 0 & 0 & 0 & 0 & 0 & 0 & 1 \\
0 & 0 & 0 & 1 & 0 & 0 & 0 & 0
\end{array}
\right)\end{array}\right),
\end{eqnarray}
where $\Gamma_{\!2}$ can then be expressed as
\begin{eqnarray}
\Gamma_{\!2}' = \left(
\begin{array}{l|lll}
A & \mbd{0} & c & \mathbf{0} \\ \hline
r & 0 & \mbd{0} & \mbd{0} \\
\mbd{0} & \mbd{0} & \mbd{0} & E \\
\mbd{0} & 1 & \mbd{0} & \mbd{0}
\end{array}
\right) \equiv \left(
\begin{array}{cc}
A & c \\
r & B \\
\end{array}
\right). \label{gamma1}
\end{eqnarray}
The reason why the fifth and sixth entries of the last line in
$\Gamma_{\!2}$ are 0 is that otherwise the rank of $\Gamma_{\!1}$
can be as large as $M$, as explained in Eq.(\ref{cm-iGamma-2}).
Further simplification can be proceeded according to the vector(or
submatrices) $c,r$. There are four cases in general, i.e., (1),
$(c=0,r=0)$; (2), $(c\neq 0,r=0)$; (3), $(c=0,r\neq 0)$; (4),
$(c\neq 0,r\neq 0)$. Here $c\neq 0$ means that $r(c)\geq 1$ and
different ranks will result in different classes, i.e.,
\begin{eqnarray}
\Gamma_{\!2}^{00} = \left(
\begin{array}{llllllll}
\times & \times & \times & 0 & 0 & 0 & 0 & 0 \\
\times & \times & \times & 0 & 0 & 0 & 0 & 0 \\
\times & \times & \times & 0 & 0 & 0 & 0 & 0 \\
0 & 0 & 0 & 0 & 0 & 0 & 0 & 0 \\
0 & 0 & 0 & 0 & 0 & 0 & 1 & 0 \\
0 & 0 & 0 & 0 & 0 & 0 & 0 & 1 \\
0 & 0 & 0 & 1 & 0 & 0 & 0 & 0
\end{array}
\right),\; _{1}\Gamma_{\!2}^{10} = \left(
\begin{array}{llllllll}
\times & \times & \times & 0 & 0 & 0 & 0 & 0 \\
\times & \times & \times & 0 & 0 & 0 & 0 & 0 \\
0 & 0 & 0 & 0 & 0 & 1 & 0 & 0 \\
0 & 0 & 0 & 0 & 0 & 0 & 0 & 0 \\
0 & 0 & 0 & 0 & 0 & 0 & 1 & 0 \\
0 & 0 & 0 & 0 & 0 & 0 & 0 & 1 \\
0 & 0 & 0 & 1 & 0 & 0 & 0 & 0
\end{array}
\right),
\end{eqnarray}

\begin{eqnarray}
\Gamma_{\!2}^{01} = \left(
\begin{array}{llllllll}
\times & \times & 0 & 0 & 0 & 0 & 0 & 0 \\
\times & \times & 0 & 0 & 0 & 0 & 0 & 0 \\
\times & \times & 0 & 0 & 0 & 0 & 0 & 0 \\
0 & 0 & 1 & 0 & 0 & 0 & 0 & 0 \\
0 & 0 & 0 & 0 & 0 & 0 & 1 & 0 \\
0 & 0 & 0 & 0 & 0 & 0 & 0 & 1 \\
0 & 0 & 0 & 1 & 0 & 0 & 0 & 0
\end{array}
\right),\; _{1}\Gamma_{\!2}^{11} = \left(
\begin{array}{llllllll}
\times & 0 & \times & 0 & 0 & 0 & 0 & 0 \\
\times & 0 & 0 & 0 & 0 & 0 & 0 & 0 \\
0 & 0 & 0 & 0 & 0 & 1 & 0 & 0 \\
0 & 1 & 0 & 0 & 0 & 0 & 0 & 0 \\
0 & 0 & 0 & 0 & 0 & 0 & 1 & 0 \\
0 & 0 & 0 & 0 & 0 & 0 & 0 & 1 \\
0 & 0 & 0 & 1 & 0 & 0 & 0 & 0
\end{array}
\right),
\end{eqnarray}

\begin{eqnarray}
_{2}\Gamma_{\!2}^{10} = \left(
\begin{array}{llllllll}
\times & \times & \times & 0 & 0 & 0 & 0 & 0 \\
0 & 0 & 0 & 0 & 1 & 0 & 0 & 0 \\
0 & 0 & 0 & 0 & 0 & 1 & 0 & 0 \\
0 & 0 & 0 & 0 & 0 & 0 & 0 & 0 \\
0 & 0 & 0 & 0 & 0 & 0 & 1 & 0 \\
0 & 0 & 0 & 0 & 0 & 0 & 0 & 1 \\
0 & 0 & 0 & 1 & 0 & 0 & 0 & 0
\end{array}
\right),\; _{2}\Gamma_{\!2}^{11} = \left(
\begin{array}{llllllll}
0 & 0 & 0 & 0 & 0 & 0 & 0 & 0 \\
0 & 0 & 0 & 0 & 1 & 0 & 0 & 0 \\
0 & 0 & 0 & 0 & 0 & 1 & 0 & 0 \\
1 & 0 & 0 & 0 & 0 & 0 & 0 & 0 \\
0 & 0 & 0 & 0 & 0 & 0 & 1 & 0 \\
0 & 0 & 0 & 0 & 0 & 0 & 0 & 1 \\
0 & 0 & 0 & 1 & 0 & 0 & 0 & 0
\end{array}
\right).
\end{eqnarray}

Clearly, analogous with the set $c_{M,l}$ in Section \ref{cml}, we
can finally get the following set
\begin{eqnarray}
c_{M-1,\;l}=\{ (\Lambda, \Gamma)|\; r(\Gamma) = l; \Gamma =
\left(\begin{array}{cc} J & 0 \\0 & B
\\ \end{array} \right); \; (\Lambda,
\Gamma)\in C_{M-1,\; l} \} \; . \label{c_N-1}
\end{eqnarray}
$J$ represents the Jordan canonical form.

\begin{theorem}
$\forall$ $(\Lambda, \Gamma) \in c_{M-i,\;l}$, the set $c_{M-i,\;
l}$ is of the classification of $C_{M-i,\;l}$. $\mathrm{(i)}$
suppose two states are SLOCC equivalent, they can be transformed
into the same matrix vector $(\Lambda, \Gamma)$; $\mathrm{(ii)}$
this matrix vector is unique in $c_{M-i,\,l}$, that is suppose
$(\Lambda,\Gamma')$ is SLOCC equivalent with $(\Lambda,\Gamma)$,
then $(\Lambda, \Gamma') = (\Lambda, \Gamma)$ $(\Gamma'=\Gamma$
means $J$s are equivalent under the condition of theorem 1 in {\rm
Ref.\cite{class-nn}} and $B'=B)$.
\end{theorem}

We give a complete classification of $2\times (M+5) \times (2M+5)$
for $M=1$, i.e., $2\times 6 \times 7$ state whose classification has
not been presented in literature so far.

\noindent Classes of sets $c_{6,\,l}$ of $2\times 6 \times 7$: for
all inequivalent classes in $c_{6,\,l}$, they have the same form of
$\Gamma_{\!1}$ in the definition (\ref{set-CN})
\begin{eqnarray}
\Gamma_{\!1} = \left[
\begin{array}{lllllll}
1 & 0 & 0 & 0 & 0 & 0 & 0 \\
0 & 1 & 0 & 0 & 0 & 0 & 0 \\
0 & 0 & 1 & 0 & 0 & 0 & 0 \\
0 & 0 & 0 & 1 & 0 & 0 & 0 \\
0 & 0 & 0 & 0 & 1 & 0 & 0 \\
0 & 0 & 0 & 0 & 0 & 1 & 0
\end{array}
\right]\; .
\end{eqnarray}
So we only list the form of $\Gamma_{\!2}$s,
\begin{eqnarray}
\left[
\begin{array}{lllllll}
\times & \times & \times & \times & \times & 0 & 0 \\
\times & \times & \times & \times & \times & 0 & 0 \\
\times & \times & \times & \times & \times & 0 & 0 \\
\times & \times & \times & \times & \times & 0 & 0 \\
\times & \times & \times & \times & \times & 0 & 0 \\
0 & 0 & 0 & 0 & 0 & 0 & 1
\end{array}
\right] , \left[
\begin{array}{lllllll}
\times & \times & \times & \times & 0 & 0 & 0 \\
\times & \times & \times & \times & 0 & 0 & 0 \\
\times & \times & \times & \times & 0 & 0 & 0 \\
\times & \times & \times & \times & 0 & 0 & 0 \\
0 & 0 & 0 & 0 & 0 & 1 & 0 \\
0 & 0 & 0 & 0 & 0 & 0 & 1
\end{array}
\right], \left[
\begin{array}{lllllll}
\times & \times & \times & 0 & 0 & 0 & 0 \\
\times & \times & \times & 0 & 0 & 0 & 0 \\
\times & \times & \times & 0 & 0 & 0 & 0 \\
0 & 0 & 0 & 0 & 1 & 0 & 0 \\
0 & 0 & 0 & 0 & 0 & 1 & 0 \\
0 & 0 & 0 & 0 & 0 & 0 & 1
\end{array}
\right] ,\label{6l0}
\end{eqnarray}
\begin{eqnarray}
\left[
\begin{array}{lllllll}
\times & \times & 0 & 0 & 0 & 0 & 0 \\
\times & \times & 0 & 0 & 0 & 0 & 0 \\
0 & 0 & 0 & 1 & 0 & 0 & 0 \\
0 & 0 & 0 & 0 & 1 & 0 & 0 \\
0 & 0 & 0 & 0 & 0 & 1 & 0 \\
0 & 0 & 0 & 0 & 0 & 0 & 1
\end{array}
\right], \left[
\begin{array}{lllllll}
0 & 0 & 0 & 0 & 0 & 0 & 0 \\
0 & 0 & 1 & 0 & 0 & 0 & 0 \\
0 & 0 & 0 & 1 & 0 & 0 & 0 \\
0 & 0 & 0 & 0 & 1 & 0 & 0 \\
0 & 0 & 0 & 0 & 0 & 1 & 0 \\
0 & 0 & 0 & 0 & 0 & 0 & 1
\end{array}
\right], \left[
\begin{array}{lllllll}
0 & 1 & 0 & 0 & 0 & 0 & 0 \\
0 & 0 & 1 & 0 & 0 & 0 & 0 \\
0 & 0 & 0 & 1 & 0 & 0 & 0 \\
0 & 0 & 0 & 0 & 1 & 0 & 0 \\
0 & 0 & 0 & 0 & 0 & 1 & 0 \\
0 & 0 & 0 & 0 & 0 & 0 & 1
\end{array}
\right] .\label{6l1}
\end{eqnarray}
Here the square matrices of $\{\times\}_{n\times n}$ in
Eq.(\ref{6l0}, \ref{6l1}) consists of all the inequivalent classes
of sets $c_{n,l}$ in $2\times n\times n$ states. For example the
first matrix of Eq.(\ref{6l0}) is made up by all the genuine
entanglement classes of the sets $c_{5,l}$ in $2\times 5\times 5$
state and plus the one with $\{\times \}_{5\times 5} = 0$, thus
there are $(26+1)$ \cite{255class} inequivalent forms of this
matrix.

\noindent Classes of set $c_{5,\,l}$ of $2\times 6 \times 7$: for
all inequivalent classes in $c_{5,\,l}$, they has the same form of
$\Lambda$
\begin{eqnarray}
\Lambda = \left[
\begin{array}{lllllll}
1 & 0 & 0 & 0 & 0 & 0 & 0 \\
0 & 1 & 0 & 0 & 0 & 0 & 0 \\
0 & 0 & 1 & 0 & 0 & 0 & 0 \\
0 & 0 & 0 & 1 & 0 & 0 & 0 \\
0 & 0 & 0 & 0 & 1 & 0 & 0 \\
0 & 0 & 0 & 0 & 0 & 0 & 0
\end{array}
\right]\; .
\end{eqnarray}
The different $\Gamma_{\!2}$s are
\begin{eqnarray}
\left[
\begin{array}{lllllll}
\times & \times & 0 & 0 & 0 & 0 & 0 \\
\times & \times & 0 & 0 & 0 & 0 & 0 \\
0 & 0 & 0 & 0 & 0 & 0 & 0 \\
0 & 0 & 0 & 0 & 0 & 1 & 0 \\
0 & 0 & 0 & 0 & 0 & 0 & 1 \\
0 & 0 & 1 & 0 & 0 & 0 & 0
\end{array}
\right], \left[
\begin{array}{lllllll}
0 & 0 & 0 & 0 & 0 & 0 & 0 \\
0 & 0 & 0 & 0 & 1 & 0 & 0 \\
0 & 0 & 0 & 0 & 0 & 0 & 0 \\
0 & 0 & 0 & 0 & 0 & 1 & 0 \\
0 & 0 & 0 & 0 & 0 & 0 & 1 \\
0 & 0 & 1 & 0 & 0 & 0 & 0
\end{array}
\right], \left[
\begin{array}{lllllll}
0 & 1 & 0 & 0 & 0 & 0 & 0 \\
0 & 0 & 0 & 0 & 1 & 0 & 0 \\
0 & 0 & 0 & 0 & 0 & 0 & 0 \\
0 & 0 & 0 & 0 & 0 & 1 & 0 \\
0 & 0 & 0 & 0 & 0 & 0 & 1 \\
0 & 0 & 1 & 0 & 0 & 0 & 0
\end{array}
\right],
\end{eqnarray}
\begin{eqnarray}
\left[
\begin{array}{lllllll}
0 & 0 & 0 & 0 & 0 & 0 & 0 \\
0 & 0 & 0 & 0 & 0 & 0 & 0 \\
0 & 1 & 0 & 0 & 0 & 0 & 0 \\
0 & 0 & 0 & 0 & 0 & 1 & 0 \\
0 & 0 & 0 & 0 & 0 & 0 & 1 \\
0 & 0 & 1 & 0 & 0 & 0 & 0
\end{array}
\right], \left[
\begin{array}{lllllll}
0 & 0 & 0 & 0 & 0 & 0 & 0 \\
0 & 0 & 0 & 0 & 1 & 0 & 0 \\
1 & 0 & 0 & 0 & 0 & 0 & 0 \\
0 & 0 & 0 & 0 & 0 & 1 & 0 \\
0 & 0 & 0 & 0 & 0 & 0 & 1 \\
0 & 0 & 1 & 0 & 0 & 0 & 0
\end{array}
\right], \left[
\begin{array}{lllllll}
0 & 0 & 0 & 1 & 0 & 0 & 0 \\
0 & 0 & 0 & 0 & 1 & 0 & 0 \\
0 & 0 & 0 & 0 & 0 & 0 & 0 \\
0 & 0 & 0 & 0 & 0 & 1 & 0 \\
0 & 0 & 0 & 0 & 0 & 0 & 1 \\
0 & 0 & 1 & 0 & 0 & 0 & 0
\end{array}
\right].
\end{eqnarray}
\begin{eqnarray}
\left[
\begin{array}{lllllll}
0 & 0 & 0 & 0 & 0 & 0 & 0 \\
1 & 0 & 0 & 0 & 0 & 0 & 0 \\
0 & 1 & 0 & 0 & 0 & 0 & 0 \\
0 & 0 & 0 & 0 & 0 & 1 & 0 \\
0 & 0 & 0 & 0 & 0 & 0 & 1 \\
0 & 0 & 1 & 0 & 0 & 0 & 0
\end{array}
\right] \; ,
\end{eqnarray}
Same as that of $c_{6,\,l}$, $\{\times \}_{2\times 2}$ here has
three different forms.

According to Proposition \ref{proposition2}, there are no true
entangled states in $C_{n,l}$ with $n<\frac{6+7}{3}$. Thus we get
$(26+1) + (13+1) + (5+1) + (2+1) + 1 + 1 +( 2+1) + 1+1+1+1+1+1= 61$
inequivalent entanglement classes in $2\times 6\times 7$. It is
clearly to see that this method is simple and effective, meanwhile
each entangled state can be read out directly from the matrix pairs.

\section{Conclusions}

In summary, we have generalized our method of entanglement
classification under SLOCC to the more general case of $2\times
M\times N$ systems. Two examples of $2\times 4\times 6$ and $2\times
6\times 7$ are given where all their inequivalent entanglement
classes are determined. Because the classification procedure is
essentially a constructive algorithm, the method can serve as a
powerful tool in practical entanglement classifications with the aid
of computers. Most importantly a wide range of state space is
explored which provide a rich resource for possible new applications
in the quantum information theory.

\vspace{.7cm} {\bf Acknowledgments} \vspace{.3cm}

This work was supported in part by the National Natural Science
Foundation of China(NSFC) under the grants 10935012, 10928510,
10821063 and 10775179, by the CAS Key Projects KJCX2-yw-N29 and
H92A0200S2, and by the Scientific Research Fund of GUCAS.

\vspace{1.5cm}

\appendix{\bf\Large Appendix}

\section{Proof of Lemma \ref{lemma1}}\label{appdixlemma1}

\noindent {\bf Necessity}: If $\mathrm{Det}(\rho_{\psi_2})=0$, there
will be ILOs that transform $\rho_{\psi_2}$ to $\rho_{\psi_2}'$ who
has at least one column and one row of zeros. Without loss of
generalities suppose the $k$th column of $\rho_{\psi_2}'$ are zeros,
for the element $(k,k)$ of $\rho_{\psi_2}'$. From
Eq.(\ref{density-cube}), we have
\begin{eqnarray}
(\rho_{\psi_2}')_{kk} & = & \sum_{ij}|(\Gamma_{i}')_{jk}|^2 = 0 \; ,
\nonumber
\end{eqnarray}
which indicates that $(\Gamma_{i}')_{jk}=0$ for all $i$ and $j$. So
the $k$th plane perpendicular to partite $\psi_2$ are all zeros.

\noindent{\bf Sufficiency}:  Suppose the $k$th ($k\leq N$) plane
perpendicular to $\psi_2$ can be transformed into a zero plane
by ILOs: $T,P,Q$. That is $\begin{pmatrix}\Gamma_1' \\
\Gamma_2'\end{pmatrix} = T\begin{pmatrix}P\Gamma_1Q \\ P\Gamma_2Q
\end{pmatrix}$, where $\Gamma'_{ijk}=0, (j=1,2\cdots M, i=1,2)$.
Then the $k$th row and $k$th column of
$[(\Gamma'_{i})^{\dag}\Gamma'_{i}]$ are all zeroes. Thus reduce
density matrix of $\rho_{\psi_2}$: $\rho_{\psi_2} =
\sum_{i}(\Gamma'_{i})^{\dag}\Gamma_{i}$, have the $k$th row and
$k$th column both zeros. So Det$(\rho_{\psi_2})=0$.

The similar proofs can be applied to $\psi_0$ and $\psi_1$, then we
have Lemma \ref{lemma1}.

\section{Proof of Proposition
\ref{proposition1}}\label{appdixproposition1}

First we prove that, in the subsets $C_{n,l}$, $\exists$ $O \in
\mathrm{ILO}$, $ O
\begin{pmatrix} \Gamma_{\!1} \\ \Gamma_{\!2} \end{pmatrix} =
\begin{pmatrix} O_{11} & O_{12} \\ O_{21} & O_{22}
\end{pmatrix} \begin{pmatrix} \Gamma_{\!1} \\
\Gamma_{\!2}\end{pmatrix}$ which  makes $r_{\mathrm{max}}(O_{11}
\Gamma_{\!1} + O_{12} \Gamma_{\!2}) = n, r_{\mathrm{min}}(O_{21}
\Gamma_{\!1} + O_{22} \Gamma_{\!2}) = l$.

\noindent Proof: in the definition $C_{n,\;l}=\{
(\Gamma_{\!1},\Gamma_{\!2})|\; r_{\mathrm{max}}(\alpha_{1}
\Gamma_{\!1} + \beta_{1} \Gamma_{\!2}) = n,
r_{\mathrm{min}}(\alpha_{2} \Gamma_{\!1} + \beta_{2} \Gamma_{\!2}) =
l \}$, (i) if Det$\begin{pmatrix} \alpha_{1} & \beta_{1} \\
\alpha_{2} & \beta_{2}
\end{pmatrix}\neq 0$, then $O = \begin{pmatrix} \alpha_{1} & \beta_{1} \\
\alpha_{2} & \beta_{2}
\end{pmatrix}$ is an ILO. (ii) if Det$\begin{pmatrix} \alpha_{1} & \beta_{1} \\
\alpha_{2} & \beta_{2}
\end{pmatrix} = 0$, then the two vectors $(\alpha_1,\beta_1),
(\alpha_2,\beta_2)$ are linearly dependent. This implies
$r(\Gamma_1)=r(\Gamma_2) = n =l$, in this case we can set
$O = \begin{pmatrix} 1 & 0 \\
0 & 1
\end{pmatrix}$.

\noindent Proof of Proposition \ref{proposition1}: Because
$\begin{pmatrix}\Gamma_1 \\ \Gamma_2\end{pmatrix} \in C_{n,l}$, so
$\exists$ $O\in \mathrm{ILO}$ that $r_{\mathrm{max}}(O_{11}
\Gamma_{\!1} + O_{12} \Gamma_{\!2}) = n, r_{\mathrm{min}}(O_{21}
\Gamma_{\!1} + O_{22} \Gamma_{\!2}) = l$. Then from $\begin{pmatrix}\Gamma_1 \\
\Gamma_2
\end{pmatrix} = T \begin{pmatrix}P\Gamma'_1Q \\ P\Gamma'_2Q
\end{pmatrix}$ we gave
\begin{eqnarray}
r_{\mathrm{max}}(O'_{11} P\Gamma'_{\!1}Q + O'_{12} P\Gamma'_{\!2}Q)
= n, \nonumber \\  r_{\mathrm{min}}(O'_{21} P\Gamma'_{\!1}Q +
O'_{22} P\Gamma'_{\!2}Q) = l \; ,
\end{eqnarray}
where $O'=OT$, $T, P, Q$ are ILOs, so we get $(\Gamma'_1, \Gamma'_2)
\in C_{n,l}$.

\section{The proof of Eq.(\ref{t-trans})} \label{TtoPQ}

$\Gamma_{\!2}$ in $(\Gamma_{\!1},\Gamma_{\!2}) \in c_{M,\,l}$ has a
form of direct sum of $J$ and $E_2$ as shown in the definition
(\ref{set-CN}). Thus when the dimension of $J$ does not equal zero,
there are no zeroes in pivot of $T'$ and the left hand side of
Eq.(\ref{t-trans}) can be separated into two parts
\begin{eqnarray}
\begin{pmatrix}
    1 & 0 \\
    \lambda & 1 \\
\end{pmatrix}
\begin{pmatrix}
    \alpha & \beta \\
    0 & \gamma
\end{pmatrix} \left(
\begin{array}{l}
E \\
J
\end{array}
\right)\; , \\
\begin{pmatrix}
    1 & 0 \\
    \lambda & 1 \\
\end{pmatrix}
\begin{pmatrix}
    \alpha & \beta \\
    0 & \gamma
\end{pmatrix} \left(
\begin{array}{l}
E_1 \\
E_2
\end{array}
\right) \; ,
\end{eqnarray}
where $\begin{pmatrix}
    1 & 0 \\
    \lambda & 1 \end{pmatrix} \begin{pmatrix}
    \alpha & \beta \\
    0 & \gamma \end{pmatrix}$ is the LU decomposition of $T'$
\cite{matrix-analysis}; $E_1$ has the same dimension as $E_2$.

For the $J$ sub-matrix we have proved \cite{class-nn} there exists
$P_{J}, Q_{J}$ which make
\begin{eqnarray}
\left(
  \begin{array}{cc}
    1 & 0 \\
    \lambda & 1 \\
  \end{array}
\right) \left(
\begin{array}{cc}
    \alpha & \beta \\
    0 & \gamma \\
  \end{array}
\right) \left(
\begin{array}{l}
P_{J} E Q_{J} \\
P_{J} J Q_{J}
\end{array}
\right) = \left(\begin{array}{l}
E \\
J
\end{array}\right) \label{J-part} \; ,
\end{eqnarray}
For the $E_{1,2}$ parts, there exist operators that
\begin{eqnarray}
P_{y} \left(\begin{array}{l} E_1 +\lambda' E_2 \\ E_2
\end{array}\right) Q_{y} = \left(\begin{array}{l} E_1  \\ E_2
\end{array}\right) \; , \nonumber
\\ P_{x} \left(\begin{array}{l} E_1  \\ E_2 + \lambda E_1
\end{array}\right) Q_{x} = \left(\begin{array}{l} E_1  \\ E_2
\end{array}\right) \; ,
\end{eqnarray}
where $\lambda'=\frac{\beta}{\alpha}$. It is simple to verify that
such kind of $P_{x,y},Q_{x,y}$ satisfying the equations does exist
(see Appendixes of \cite{class-nn} for detailed derivations). Thus
$P_{C}=P_{x}P_{y}$ and $Q_{C}=Q_{y}Q_{x}$ will make
\begin{eqnarray}
\left(
\begin{array}{cc}
    1 & 0 \\
    \lambda & 1 \\
  \end{array}
\right) \left(
  \begin{array}{cc}
    \alpha & \beta \\
    0 & \gamma \\
  \end{array}
\right) \left(
\begin{array}{l}
P_{C} E_1 Q_{C} \\
P_{C} E_2 Q_{C}
\end{array}
\right) = \left(\begin{array}{l}
E_1 \\
E_2
\end{array}
\right) \label{C-part} \; .
\end{eqnarray}
Combine Eq.(\ref{J-part}) and Eq.(\ref{C-part}) we can get such
$P_{0} = P_{J}\oplus P_{C}$, $Q_{0} = Q_{J}\oplus Q_{C}$ that
satisfy the following equation
\begin{eqnarray}
\begin{pmatrix}
    1 & 0 \\
    \lambda & 1 \\
\end{pmatrix}
\begin{pmatrix}
    \alpha & \beta \\
    0 & \gamma
\end{pmatrix} \left(
\begin{array}{l}
P_{0}\Gamma_{\!1}Q_{0} \\
P_{0}\Gamma_{\!2} Q_{0}
\end{array}
\right) = \left(
\begin{array}{l}
\Gamma_{\!1} \\
\Gamma_{\!2}
\end{array}
\right)\; ,
\end{eqnarray}
which is just Eq.(\ref{t-trans}).

However there exists the special case that the dimension of $J$
equals zero, in this case there can be zero elements in the pivot of
the nonsingular square matrix $T'$. $T'$ can then be decomposed as
decomposed as \cite{matrix-analysis}
\begin{eqnarray}
\begin{pmatrix}
    t_{11}' & t_{12}' \\
    t_{21}' & t_{22}' \end{pmatrix} = P_{T'} \cdot \begin{pmatrix}
    1 & 0 \\
    \lambda & 1 \end{pmatrix} \cdot \begin{pmatrix}
    \alpha & \beta \\
    0 & \gamma \end{pmatrix} \; ,
\end{eqnarray}
where $\alpha, \beta, \gamma, \lambda \in \mathbb{C}$, $P_{T'} =
\begin{pmatrix} 0 & 1 \\ 1 & 0 \end{pmatrix}$ and both
matrices on the righthand side of above equation are nonsingular. It
can be show that $P_{T'}$ can be compensated by some operators
$P_{z}, Q_{z}$ which act on $\Gamma_{\!1}$ and $\Gamma_{\!2}$, i.e.,
\begin{eqnarray}
\left(\begin{array}{l} E_1 \\ E_2 \end{array}\right) = P_{T}
\left(\begin{array}{l} P_{z} E_1 Q_{z}\\ P_{z} E_2 Q_{z}
\end{array}\right)\; , \label{permut}
\end{eqnarray} see Appendixes of \cite{class-nn}.

\section{Proof of Eq.(\ref{cm-iGamma-2})}\label{App-C}

\noindent First we prove the following proposition.

$\forall$ $(\Gamma_1,\Gamma_2) \in C_{M-i,l}$ and $\alpha,\beta \neq
0$: if (1) $\Gamma_1$ has the form of Eq.(\ref{Gamma1-S-Form}) and
$\Gamma_2$ has the following structure
\begin{eqnarray}
\Gamma_{\!2} = \left(\begin{array}{l|l|l|l|l} \left[\begin{array}{ll}\times & \times \\
\times & \times
\end{array}\right]_{k\times k} & \begin{array}{l} \times  \\ \times \end{array} &
\begin{array}{l} \times \\ \times \end{array}
& \begin{array}{l} \mbd{0} \\ \mbd{0} \end{array} & \begin{array}{l} \times \\ \times \end{array} \\
\hline \begin{array}{lr} \mbd{0} & \ \mbd{0} \end{array} & \
\mbd{0}_{I\times I} & \ \mbd{0}
& E_{I\times I} & \ \mbd{0} \\ \hline \begin{array}{ll} \times & \times \end{array} & \ X & \ \mbd{0} & \ \mbd{0} & \ \mbd{0} \\
\begin{array}{ll}  \mbd{0} & \ \mbd{0} \end{array} & \ \mbd{0} & \ \times & \ \mbd{0} & \ \times \\
\begin{array}{ll}  \mbd{0} & \ \mbd{0} \end{array} & \ \mbd{0} & \ \times & \ \mbd{0}  &
\ \times
\end{array}\right) \; , \label{app-C-1}
\end{eqnarray}
where $\Gamma_2(\{ k+1,\cdots ,k+I\}, \{l+1,\cdots,l+I\}) =
E_{I\times I}$; (2) $\Gamma=\alpha\Gamma_1 + \beta\Gamma_2$,
$r(\Gamma(\{ k+I+1,\cdots,M \},\{ k+I+1,\cdots, N \})) =
r(\Gamma_1(\{ k+I+1,\cdots,M \},\{ k+I+1,\cdots, N \})) = M-i-k-I$.
Then the rank $r(\Gamma(\{ k+1,\cdots,M \},\{ k+1,\cdots, N \}))$ is
larger when $X \neq 0$ than $X =0$.

\noindent {\bf Proof:} Because $r(\Gamma(\{ k+I+1,\cdots,M \},\{
k+I+1,\cdots, N \})) = r(\Gamma_1(\{ k+I+1,\cdots,M \},\{
k+I+1,\cdots, N \})) = M-i-k-I$ then the column vectors of
$\Gamma_2(\{ k+I+1,\cdots,M \},\{ k+I+1,\cdots, N \})$ are linearly
dependent on that of $\Gamma_1$. From Eq.(\ref{app-C-1}) we have if
$X=0$ then $r(\Gamma(\{ k+1,\cdots,M \},\{ k+1,\cdots, N \})) =
M-i-k$; if $X\neq 0$ then
\begin{eqnarray}
r(\Gamma(\{ k+1,\cdots,M \},\{ k+1,\cdots, N \})) = M-i-k+r(X)
>M-i-k \nonumber \; ,
\end{eqnarray}
which complete the proof.

This indicates that if we insist the maximum rank of
$r(\Gamma=\alpha\Gamma_1+\beta\Gamma_2)=M-i$ then $X=0$. Thus
$\forall (\Gamma_1, \Gamma_2)\in C_{M-i,l}$, we have $X=0$.
Similarly argument applies to $[\Gamma_2(\{1, \cdots, M\},\{
1,\cdots, M-i \})]^{T}$. Then we can get Eq.(\ref{cm-iGamma-2}).

\end{document}